\documentclass[twocolumn]{aa}  
\usepackage{linenoaa}
\usepackage{amsmath,amssymb,amsfonts}
\usepackage{graphicx}
\usepackage{xcolor}
\usepackage{color}
\usepackage{txfonts}

\usepackage{hyperref}
\hypersetup{
    colorlinks=true,
    linkcolor=blue,
    citecolor=blue,
    filecolor=magenta,      
    urlcolor=blue,
    pdftitle={The impact of circumstellar WBBs on the dynamics of PWNe and SNRs}
    }

\begin{document} 
\title{The impact of circumstellar wind-blown bubbles on the dynamics of pulsar wind nebulae and supernova remnants}

\author{Lioni-Moana Bourguinat\inst{1,2}\thanks{lioni-moana.bourguinat@gssi.it}
\and Pierrick Martin\inst{3}
}

\institute{
Gran Sasso Science Institute (GSSI), Viale Francesco Crispi 7, 67100 L’Aquila (AQ), Italy
\and
INFN-Laboratori Nazionali del Gran Sasso (LNGS), via G. Acitelli 22, 67100 Assergi (AQ), Italy
\and
IRAP, Université de Toulouse, CNRS, CNES, F-31028 Toulouse, France
}

\date{Received Month XX, XXXX; accepted Month XX, XXXX}

  \abstract
   {The core-collapse supernova explosion of most massive stars is followed by the development of a supernova remnant (SNR) and pulsar wind nebula (PWN) system whose evolution is in part dictated by its surrounding environment. The latter is frequently assumed to be the typical interstellar medium (ISM). However, massive stars emit powerful winds during their lifetime and carve extended circumstellar cavities known as wind-blown bubbles (WBBs).}
   {We investigated the impact of WBBs on the dynamics, energetics, and thermal X-ray emission of a SNR-PWN system.}
   {We performed numerical simulations of a 1D hydrodynamical analogue of the system, for different stellar progenitors and pulsar parameters, in a WBB or ISM environment.}
   {The system initially expands to large sizes in the low-density interior of a WBB, until the SNR fills the entire bubble. The PWN grows to $\gtrsim10~\mathrm{pc}$ ($30~\mathrm{pc}$) after $\sim5~\mathrm{kyr}$ ($\sim25~\mathrm{kyr}$) for a $M=8~\mathrm{M}_\odot$ ($M=20~\mathrm{M}_\odot$) progenitor, $5-6$ times larger than in an ISM environment. The impact of the SNR on the massive outer shell of the WBB generates a strong reflected shock that efficiently returns energy to the inner parts of the system. At late times, after a strong compression of the PWN by an order of magnitude in size, back-and-forth shock crossings of the cavity maintain the SNR at very high temperatures and drive a vigorous reverberation of the PWN, with changes in radius by up to $3-4$. In this stage, the PWN is on average $2-4$ times larger than in an ISM environment and contains up to an order of magnitude more energy. Thermal radiation from the system occurs in few very localised emission bursts, as low-velocity shocks are transmitted across the bubble shell when it is hit by blast waves.}
   {WBBs offer a possible explanation for the large sizes of many PWNe and the non-detection of their parent SNRs.}


\titlerunning{The impact of circumstellar WBBs on the dynamics of PWNe and SNRs}

\maketitle
\nolinenumbers

\section{Introduction}

At the end of their lives, massive stars explode as core-collapse supernovae, ejecting their stellar envelope at supersonic speeds, while in the majority of cases the core collapses into a fast rotating and highly magnetised neutron star called a pulsar \citep{sukhbold+2016}. The stellar ejecta interacts with the environment to form a supernova remnant (SNR), in which the kinetic energy of the explosion is dissipated into thermal energy \citep{dwarkadas2005}. Meanwhile, the rotational energy of the pulsar is progressively lost to an ultra-relativistic magnetised outflow loaded with energetic leptons produced in the rotating magnetosphere and escaping it along open field lines. The expansion of this wind is halted at a termination shock (TS), where the dynamic pressure of the wind matches the static pressure of the external environment, which is the supernova ejecta in the first $\sim 10-100~\mathrm{kyr}$ of the evolution of the system \citep{gaensler+2006}. At or close to the TS, the (re)acceleration of the leptons creates a population of non-thermal particles that spread downstream into a hot bubble growing inside the SNR and forming a pulsar wind nebula or PWN \citep{chevalier+1992}.

The non-thermal particles emit synchrotron and inverse Compton radiation that can be observed from radio wavelengths to X-ray and gamma-ray energies. In the latter domain, decades after the identification of the Crab nebula as the first observed PWN above $100~\mathrm{GeV}$ \citep{weekes+1989}, their number has grown so much that they constitute the largest class of identified sources in the most recent H.E.S.S., HAWC, and LHAASO catalogues \citep{HESS2018, HAWC2026, LHAASO2024}. This seems to hold in recent observations at the highest photon energies, above $100~\mathrm{TeV}$, suggesting that PWNe are the most prolific and efficient particle accelerators in the Galaxy \citep{Albert2021,Breuhaus2022,DeOnaWilhelmi2022}. As common gamma-ray sources and unrivalled laboratories to investigate the acceleration of cosmic-ray leptons, PWNe have been studied extensively. Various modelling approaches were used to account for the different morphological and spectral features of the PWNe sources \citep[see][for a comprehensive overview]{olmi+2023}.

Despite many significant advancements in the understanding of PWNe over the decades, some discrepancies between observations and simulations persist. For this study we  focused on two in particular. Firstly, several works point to the fact that many observed PWNe seem to exhibit larger extents than what could be expected from reasonable assumptions. This is true in particular for samples of sources detected in X-rays \citep{bamba+2010} and/or gamma-rays \citep{HESS2018b}. Physical sizes of a few tens of parsecs are inferred, at ages up to $100~\mathrm{kyr}$. This is inconsistent with the expectation that, after a phase of free expansion over the first millenaries, the PWN will be compressed by the inward-going reverse shock of the SNR and its subsequent evolution will be constrained by the pressurised remnant \citep{bandiera+2023a}. A proposed explanation to this size discrepancy invokes a high degree of particle escape across the system: non-thermal leptons would escape into the parent SNR and out to the surrounding medium and radiate from there, especially in inverse-Compton gamma-rays \citep[see][and references therein]{martin+2024,martin+2025}. Secondly, around several PWNe with ages below $100~\mathrm{kyr}$, the SNR is either barely detectable or not found at all. This is true in particular for the iconic Crab nebula. In this case, the anomalously high luminosity of the historical supernova could be the result of a relatively low-mass progenitor whose explosion happened inside a specific circumstellar medium (CSM) containing up to $3~\mathrm{M}_\odot$ in a $10^{15}~\mathrm{cm}$ radius \citep{smith2013}, a scenario supported by \citet{temim+2024}. After an early interaction of the SNR with this CSM, what we witness now is the PWN bursting out of the cocoon formed. Meanwhile, the search for a larger shell around the Crab nebula in X-rays has been unsuccessful \citep{frail+1995,seward+2006}. The search for the missing remnants has been going on for more than two decades \citep[see][and references therein for some examples]{matheson+2010}, and progress has been made from deep X-ray and radio searches. Nevertheless, some emblematic very extended and gamma-ray bright PWNe still lack a clearly identified parent SNR and this hinders a solid interpretation of their properties: HESS J1825-137 \citep{Abdalla2019}, HESS J1809-193 \citep{Aharonian2023}, or HESS J1813-178 \citep{Aharonian2024}, for instance. In other cases, emission structures vaguely suggestive of a SNR are observed, but their relation to non-thermal emission from the associated pulsar and PWN are ambiguous and not in line with a classical PWN+SNR picture, for instance SNR G106.3+2.7, which is an ultra-high-energy source and major PeVatron candidate \citep{MAGIC2023,DeSarkar2022}.

A possible solution to both problems may arise from the wind-blown bubbles (WBBs) that massive star progenitors produce around them. The powerful winds of these stars modify the surrounding interstellar medium (ISM) into a stratified environment with peculiar density, pressure, and temperature profiles \citep{weaver+1977}. In a set-up similar to PWNe, supersonic stellar winds thermalise at a termination shock and feed a hot high-pressure and low-density plasma, extending up to several tens of parsecs (after a few megayears) and driving away a massive gas shell comprising all the interstellar material that was in place at the onset. The tenuous interior of the WBB is made up of shocked stellar wind material and matter evaporated from the cold and dense shell that marks the outer boundary between the WBB and the ISM. 
The evolution of SNRs inside WBBs was explored with self-similar solutions by \citet{chevalier+1989} and semi-analytical or fully numerical solutions by \citet{tenorio-tagle+1990} and \citet{dwarkadas2005}.

From the point of view of observations, the clear identification of a SNR expanding inside a WBB can be quite challenging, and traditional probes most likely favour the detection of SNRs evolving in a sufficiently dense medium \citep{chu1997}, either the undisturbed ISM -- if the wind bubble was small and rapidly overtaken by the SNR or if the progenitor star managed to escape the bubble blown during main sequence owing to its proper motion -- or the unshocked stellar wind of the progenitor star -- if the SNR is still propagating in the innermost regions of the bubble. The presence of WBBs is often envisioned through their indirect impact on some observational properties of SNRs.
For instance, the presence of radiative shocks at a very early age in the Cygnus Loop is hard to explain for a SNR developing in the normal ISM, but makes sense assuming an asymmetrical SNR expanding at various speeds in a WBB \citep[and references therein, as well as other examples]{vink2012}.
As another example, \citet{chiotellis+2024} showed that mixed-morphology SNRs (also called thermal composite SNRs), that exhibit bright X-ray cores and radio shells, could be explained as SNRs expanding in WBBs \citep[see the list of objects in][]{zhang+2015}.
Finally, \citet{khabibullin+2024} explains the age discrepancy of the apparently old S147 SNR, associated with the much younger pulsar J0538+2817, as a result of its evolution inside a WBB.

However, although we are well motivated from observations and theories of massive star evolution or SNR dynamics, WBBs are not systematically considered for their indirect impact on PWNe. Instead, it is frequently assumed that PWN+SNR systems develop in a uniform medium with densities typical of the quiescent ISM. In this framework, the SNR undergoes a canonical evolution consisting of a free expansion phase, followed by the Sedov-Taylor phase, and a subsequent radiative phase \citep{truelove+1999,cioffi+1988}. The first two stages are relevant to PWNe that are younger than a few tens of kiloyears \citep{vanderswaluw+2001,bucciantini+2003}. Initially, the PWN goes through a free expansion stage in the cold stellar ejecta, until the reverse shock of the SNR progressively heats up the ejecta again and eventually collides with the PWN. This leads to strong compression of the nebula and initiates a sequence of reverberation during which the PWN oscillates in size \citep{bandiera+2020,bandiera+2023a}. In this framework, the importance of some environmental properties on the evolution of the PWN were considered, such as density gradients or anisotropies in the ISM that lead to asymmetries in the reverse shock dynamics and PWN compression, and a displacement of the nebula with respect to its pulsar \citep{temim+2009, slane+2019}.

In this paper we report the results of a series of numerical experiments conducted to assess the impact of a WBB on the dynamics of a PWN, with a particular emphasis on the extent of the nebula and the state of the remnant, in particular its ability to radiate in X-rays. Since we are primarily interested in how a realistic environment affects the confinement of the PWN, we implemented a simple hydrodynamical analogue of a PWN+SNR system and neglected complicated magnetic and relativistic phenomena that occur in the innermost regions of the system, in the close vicinity of the pulsar. In a nutshell, the two major effects we investigated here are a very low density for the medium in which the parent supernova explodes, and the presence of a massive gas wall at some distance from the progenitor star. A number of other ingredients of the problem can have significant consequences on the development of the whole system,  the ambient magnetic field strength, the progenitor star velocity, the magnitude and orientation of the pulsar kick, or an asymmetry in the WBB, to name a few. \citet{meyer+2022,meyer+2024,meyer+2024b,meyer+2025,meyer+2025b} investigated the influence of some of these for a selected set of parameter combinations through 2D or 3D magneto-hydrodynamic simulations and illustrated how complex the system can become.

The outline of this paper is as follows. We present the physical hypotheses and the numerical set-up in Sect.~\ref{sec:model}. We discuss the results of the simulations in Sect.~\ref{sec:results}. We summarise our findings in Sect.~\ref{sec:conclusions}.

\section{Model}
\label{sec:model}

We implemented a purely hydrodynamical analogue of a SNR-PWN system. In practice, it means that the actual relativistic pulsar wind carrying energetic electron-positron pairs and magnetic field is approximated by a fast outflow of cold and low-density normal monoatomic gas emanating from a central point-like source. The PWN in our simulations is a bubble of low-density and very-high-temperature normal gas expanding in ejecta composed of a similar monoatomic gas initially at much higher density and much lower temperature (see the discussion in Sect. \ref{sec:model:num} about the adiabatic index).

We perform a series of simulations of the development of a PWN for a variety of conditions. We consider two stellar progenitors with $M_\mathrm{ZAMS}=8~\mathrm{M}_\odot$ and $20~\mathrm{M}_\odot$ (referred to as M8 and M20 in the following) that are thought to bracket the range of zero-age main-sequence stellar masses yielding neutron stars, although there are large uncertainties on that matter \citep{sukhbold+2016,schneider+2021}. However, given the steep initial mass function of massive stars, such a range should capture the majority of pulsar-producing stars. Second, we consider five different pulsars with different initial spin-down powers and timescales, in order to cover the typical spread of these quantities that is obtained from pulsar population synthesis exercises. Last, we simulate the evolution of the systems in both the WBBs appropriate for each progenitor but also in a uniform ISM in order to provide a reference case for comparison.

In the following sections, we present the description adopted for each component of the systems we simulated: (i) the surrounding environment structure, WBB or ISM; (ii) the properties of the SNR propagating in this environment; (iii) the pulsar and its associated PWN developing at the centre of the system. For many assumptions on the WBB, we take up the work introduced in \citet{bourguinat+2026} and we refer to this publication for detailed formulae and hypotheses. 

\subsection{WBB}
\label{sec:model:wbb}

For the description of WBBs created by massive stars, we used the seminal picture introduced in \citet{weaver+1977}. The innermost region of the bubble is made of freely expanding stellar wind until the wind termination shock found at radius $r_\mathrm{w}$. Beyond this wind region, a cavity of density $\rho_\mathrm{b}$ and temperature $T_\mathrm{b}$ filled with the hot and tenuous material extends up to the bubble outer radius $r_\mathrm{b}$. Then, marking the boundary of the bubble with the Galactic ISM at $r_\mathrm{shell}$, one finds a cold, geometrically thin, and dense shell of parameters $\rho_\mathrm{shell}$ and $T_\mathrm{shell}$, that is composed of all the pre-existing interstellar material swept-up in the expansion of the WBB. At the time of supernova explosion, the size of the WBB is mostly set by the main-sequence winds while its inner pressure and density structure is determined by the red supergiant winds.

Table~\ref{tab:parameters_WBB} is reproduced from \citet{bourguinat+2026} and lists the main WBB parameters for the M8 and M20 progenitors. The stellar wind properties are essentially derived from stellar evolution calculations by \citet{ekstrom+2012} for non-rotating stars. We highlight a few important numbers: the bubble shells are very massive, of the order of $\sim10^3-10^4~\mathrm{M}_\odot$, which is much larger than the typical ejecta masses $\sim10~\mathrm{M}_\odot$; the bubble interiors are very tenuous, with gas densities of the order of $\sim10^{-5}-10^{-4}~\mathrm{cm}^{-3}$, which makes it almost impossible for a SNR to reach the Sedov-Taylor phase during their propagation inside the bubble; the bubble size increases from 20 to $45~\mathrm{pc}$ with progenitor mass, which will have major consequences as we show below. \citet{khabibullin+2024} computed similar shell masses for the S147 system that is assumed to lie in a $39~\mathrm{pc}$ cavity created by a $20~\mathrm{M}_\odot$ progenitor.

While the work presented here relies on a single set of parameters for each progenitor mass, we acknowledge that some variance is expected and observed and will naturally have an impact on the bubble properties. At any given stellar mass, the mass-loss rate can vary by at least an order of magnitude \citep{seo+2018,puls+2008}, while the ambient density can span four orders of magnitude, from 0.1 to 1000 H/cm$^3$ \citep[from warm ionised to cold diffuse molecular gas; see][]{ferriere2001}. Although wind-blown bubble properties have a rather weak dependence on these parameters \citep{weaver+1977}, which somewhat mitigates the problem, the large range of possible ambient densities at least will translate into very different bubble sizes and shell masses. We defer the investigation of such effects to future work, and caution for the time being that the results presented here are rather representative of systems evolving in the warm phase of the ISM.

\begin{table}
    \caption{Properties of the stellar progenitors and associated WBB environments at the time of supernova.}
    \label{tab:parameters_WBB}
    \centering
    \begin{tabular}{ccc}
    \hline \hline
    Progenitor Mass {[}M$_\odot${]}       & 8                  & 20                 \\ \hline
    MS wind luminosity {[}erg/s{]}        & $1.5\times10^{31}$ & $5.1\times10^{34}$ \\
    RSG mass loss rate {[}M$_\odot$/yr{]} & $2.9\times10^{-7}$ & $1.2\times10^{-5}$ \\
    RSG wind speed {[}km/s{]}             & 11.0               & 23.4              \\
    RSG wind radius {[}pc{]}              & 16.4               & 17.2              \\
    Bubble density {[}cm$^{-3}${]}        & $5.3\times10^{-5}$ & $5.0\times10^{-4}$ \\
    Bubble temperature {[}K{]}            & $6.2\times10^{4}$  & $5.0\times10^{5}$  \\
    Bubble radius {[}pc{]}                & 20.1               & 45.4               \\
    Shell density {[}cm$^{-3}${]}         & 4.5                & 10.6               \\ \hline
    \end{tabular}
    \tablefoot{Adapted from \citet{bourguinat+2026}.}
    
\end{table}

\subsection{Supernova remnant}
\label{sec:model:snr}

For the explosion resulting from the final gravitational collapse of our M8 and M20 progenitors, we assumed that the same kinetic energy $E_\mathrm{ej}=10^{51}~\mathrm{erg}$ is imparted to 
the ejecta. Following \citet{bourguinat+2026}, the mass of the latter is computed from the stellar evolution models used in \citet{ekstrom+2012} and amounts to $M_\mathrm{ej}=3.1~\mathrm{M}_\odot$ and $10.4~\mathrm{M}_\odot$, respectively. The initial density profile consists of a core of constant density $\rho_\mathrm{c}$ and an envelope with a power-law density of index $9$. The initial velocity profile is a linear function of radius starting from zero at the centre \citep[][and references therein]{truelove+1999}.

We assume that the supernova explosion takes place at the centre of the WBB, for simplicity as this allows to run 1D simulations. This is an important assumption, not necessarily always verified in nature as massive stars in our Galaxy are observed to have space velocities with a typical dispersion of $10~\mathrm{km/s}$ \citep{bobylev+2022}, which implies displacements comparable to the WBB sizes over their lifetimes. Such a consideration is particularly relevant to the fact that the mass in the bubble shell in our simulations is so large compared to the ejecta mass that it nearly acts as a static wall. As detailed below, this places us in the case of a very massive shell investigated by \citet{dwarkadas2005}, with as major consequence the creation of a very strong reflected shock propagating inwards that will quickly thermalise the supernova ejecta (and incidentally violently compress the PWN). An offset explosion, closer to the bubble shell, would lead to a contrasting development with limited growth and quick transition to a radiative phase on one side and a free expansion up to a large extent on the opposite side.

\subsection{PWN}
\label{sec:model:pwn}

In our simulations, the pulsar is described as the central point-like source of a high-velocity and low-density outflow. Its main parameters are its initial spin-down luminosity and its spin-down time, which determines the amount of mechanical energy progressively injected into the system and the typical duration over which this is achieved. The observed pulsar population of our Galaxy results from orders-of-magnitude variations in the initial parameters of neutron stars \citep{faucher-giguere+2006,watters+2011}, which can be expected to leave an imprint on the associated PWNe. We therefore considered several possibilities for the pulsar component so as to bracket the core of the pulsar population. The corresponding parameter sets are listed in Table~\ref{tab:pulsar_parameters}: from left to right, the pulsar identifier, the total rotational energy, the initial spin-down timescale, and the initial spin-down luminosity. Pulsars are identified as LXX.XTYY.Y, where XX.X and YY.Y are the base-10 logarithms of the initial spin-down luminosity and initial spin-down time, respectively. Pulsar model L37.5T3.6  corresponds to the most typical pulsar, lying at the mean of the distributions of both initial spin-down luminosity and spin-down time. Overall, our choices translate into an order of magnitude scatter in total rotational energy. For the time profile of the spin-down power, we assume the typical power-law decline with a pulsar braking index of $n=3$.

As detailed below, our numerical simulations compute the evolution of the system starting from a few decades past supernova explosion. The initial state of the PWN at that moment is imposed following the analytical descriptions given in \citet{blondin+2001} and \citet{vanderswaluw+2001}. For computational reasons, the pulsar outflow, normally a relativistic wind, is ascribed a speed that is twice the outer velocity of the SNR ejecta and a pressure that is $10^{-7}$ that of the ISM. We tested alternative prescriptions and found no impact on the development of the PWN on large scales. We also emphasise that our spherically symmetric numerical set-up implies a zero initial kick velocity for the pulsar. Given the typical development of the PWN in the WBB case ($5-10~\mathrm{pc}$ after $30~\mathrm{kyr}$ for M8 and $10-15~\mathrm{pc}$ after $90~\mathrm{kyr}$ for M20; see Sect. \ref{sec:results}), kick velocities below $100-150~\mathrm{km/s}$ would not change the outcome, as the pulsar would remain inside the nebula throughout, so our simulations still capture a fair fraction of the population according to the natal kick velocity distribution inferred in \citet{disberg+2025}. We further discuss the impact of kick velocities in Sect. \ref{sec:dynamics}.

\begin{table}
\centering
\caption{Pulsar parameters used in the simulations.}
\label{tab:pulsar_parameters}
\begin{tabular}{cccc}
\hline
Model & $E_0$ {[}erg{]} & $\tau_0$ {[}yr{]} & $L_0$ {[}erg/s{]} \\ \hline\hline
L37.4T3.3     & $1.6\times10^{48}$                & $2.0\times10^{3}$      & $2.5\times10^{37}$                      \\
L36.0T4.9     & $2.5\times10^{48}$                & $7.9\times10^{4}$      & $1.0\times10^{36}$                      \\
L37.5T3.6     & $4.0\times10^{48}$                & $4.0\times10^{3}$      & $3.2\times10^{37}$                      \\
L39.0T2.4     & $7.9\times10^{48}$                & $2.5\times10^{2}$      & $1.0\times10^{39}$                      \\
L37.7T4.0     & $1.6\times10^{49}$                & $1.0\times10^{4}$      & $5.0\times10^{37}$                      \\ \hline
\end{tabular}
\end{table}

\subsection{Numerical methods}
\label{sec:model:num}

We ran our hydrodynamical simulations using a Godunov-type finite-volume code called Idefix \citep{lesur+2023}. We configured Idefix so that it solves the conservation equations with a second-order-accurate piecewise-linear cell reconstruction that uses a van Leer slope limiter, and inter-cell flux computations using the Harten-Lax-van Leer-Contact solver (HLLC). The time integration is done by an explicit second order Runge-Kutta scheme. The numerical stability of the time marching is assured by a Courant-Friedrichs-Lewy condition that we fixed to $0.5$.

For each simulation, the physical quantities $x$ are reduced to a system of characteristic units depending on the main parameters \citep{truelove+1999}: $x^\star = x / x_\mathrm{ch}$. Since we use the same ejecta energy and outer density in all runs, our characteristic units $x_\mathrm{ch}$ depend only on the ejecta mass hence the progenitor mass: $r_\mathrm{ch} = 4.491~\mathrm{pc}$ and $t_\mathrm{ch} = 1096~\mathrm{yr}$ for M8, $r_\mathrm{ch} = 6.703~\mathrm{pc}$ and $t_\mathrm{ch} = 2983~\mathrm{yr}$ for M20.

Assuming spherical symmetry, we used a one-dimensional radial position grid from $r^\star=10^{-5}$ to $r^\star=8$. The grid is divided in three blocks: a first block of 2500 cells with uniform spacing until $r^\star=0.1$ so that the pulsar wind is contained in several tens of cells, a block with stretch spacing of 1500 cells until $r^\star=1$, and finally a block with uniform spacing of 6000 cells. The simulations are run from $t^\star = 0$ to $30$, and actual time is obtained as $t = t_\mathrm{ini} + t^\star\times t_\mathrm{ch}$, with $t_\mathrm{ini}=60~\mathrm{yr}$. This translates into simulation end times of $\sim35~\mathrm{kyr}$ and $\sim90~\mathrm{kyr}$, for the M8 and M20 progenitors respectively.

The simulation starts at $t_\mathrm{ini}$, in the WBB appropriate to each stellar progenitor or in the undisturbed ISM, from a situation where the stellar ejecta were allowed to expand freely and interactions with the surrounding medium were neglected, while the PWN has started its development in the cold overlying ejecta and its initial structure is based on the analytical formulae of \citet{blondin+2001}. At this time, from the outer edge of the grid to the centre, one finds (in the WBB case):
(i) the ISM at rest, with a number density of $1~\mathrm{cm}^{-3}$ and a temperature of $8000~\mathrm{K}$ \citep{recchia+2022};
(ii) the bubble shell at rest;
(iii) the interior of the bubble at rest;
(iv) the freely expanding stellar wind of the progenitor star;
(v) the freely expanding stellar ejecta;
(vi) a small shell of swept-up ejecta material;
(vii) the nebula;
(viii) the freely expanding pulsar wind.

We used a single adiabatic index of 5/3 in the entire domain, similar to works like \cite{vanderswaluw+2001} or \cite{meyer+2026} but different from others like \cite{blondin+2001} or \cite{bucciantini+2003}, where a value of 4/3 was used on account of the relativistic nature of the pulsar wind material. The impact of this choice is minor compared to the effects we are trying to illustrate here \citep[see the discussion in][]{bucciantini+2003}. In addition, as a mix of ultra-relativistic pairs and magnetic field, the actual adiabatic index of gas in the nebula may be higher than 4/3, depending on the actual state of the flow and associated magnetic turbulence \citep{mckee+1995}.

We implemented thermal radiation losses using the cooling function of \citet{schure+2009}, for a solar metallicity plasma with temperatures in the $10^4-10^8~\mathrm{K}$ range. We neglected non-thermal radiation losses, in particular synchrotron losses from the nebula, which in reality is a mixture or ultra-relativistic electron-positron pairs and magnetic field but is simply approximated as a hot gas in our simulations. This amounts to overestimating the internal energy and radial extent of the PWN, at any time but particularly after the first compression that causes a surge of synchrotron losses owing to the strong increase of the nebular magnetic field strength \citep{bandiera+2023b}.

We validated our approach by checking numerical solutions against analytical expectations for both the SNR and PWN dynamics in the simple case of a propagation in a uniform ISM. Our numerical results for the SNR evolution were compared to the analytical formulae from \citet{truelove+1999} and showed excellent agreement for the forward shock (FS), at least until radiative losses become significant, and a satisfactory agreement for the reverse shock (RS), with deviations resulting from approximations in the formulae and consistent with those already apparent in \citet{truelove+1999}. The PWN dynamics was compared to Eq.~12 and Eq.~21 of \citet{vanderswaluw+2001} for the early and late evolutions the nebula, respectively. Under the assumption of a constant-luminosity pulsar, the early evolution assumes a PWN expanding in the cold ejecta of a freely expanding SNR, while the late evolution assumes a PWN growing in the pressurised ejecta of a Sedov-Taylor SNR. For a meaningful test, we limited the comparison to pulsar model L36.0T4.9, which has the longest spin-down time of $79~\mathrm{kyr}$. We found perfect agreement between our numerical results and the analytical solutions for the early evolution, while the dynamics asymptotically tends to the trend predicted by \citet{vanderswaluw+2001} at late times, as oscillations from the reverberation phase are progressively damped.

\section{Results}
\label{sec:results}

In this section, we discuss the outcome of our numerical simulations, focusing primarily on four set-ups: the PWN powered by pulsar model L37.5T3.6, developing inside the SNR from either of our two M8 and M20 stellar progenitors, which expands in a uniform ISM or in a stratified WBB. We  illustrate the impact of using different pulsar properties for a subset of the results, in particular the extent of the PWN.

\subsection{Dynamics}
\label{sec:dynamics}

\begin{figure}
    \centering
    \includegraphics[width=0.49\textwidth]{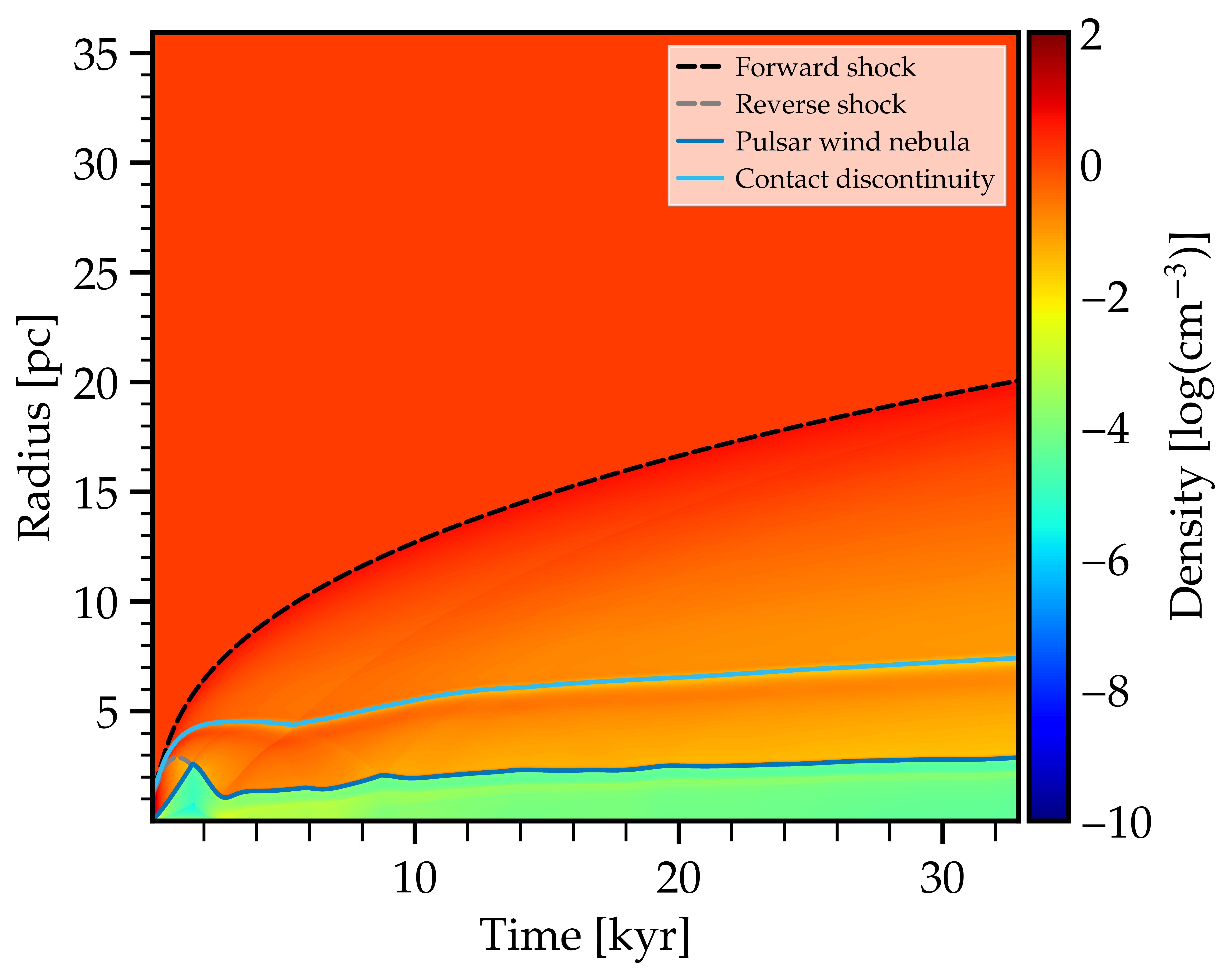}
    \includegraphics[width=0.49\textwidth]{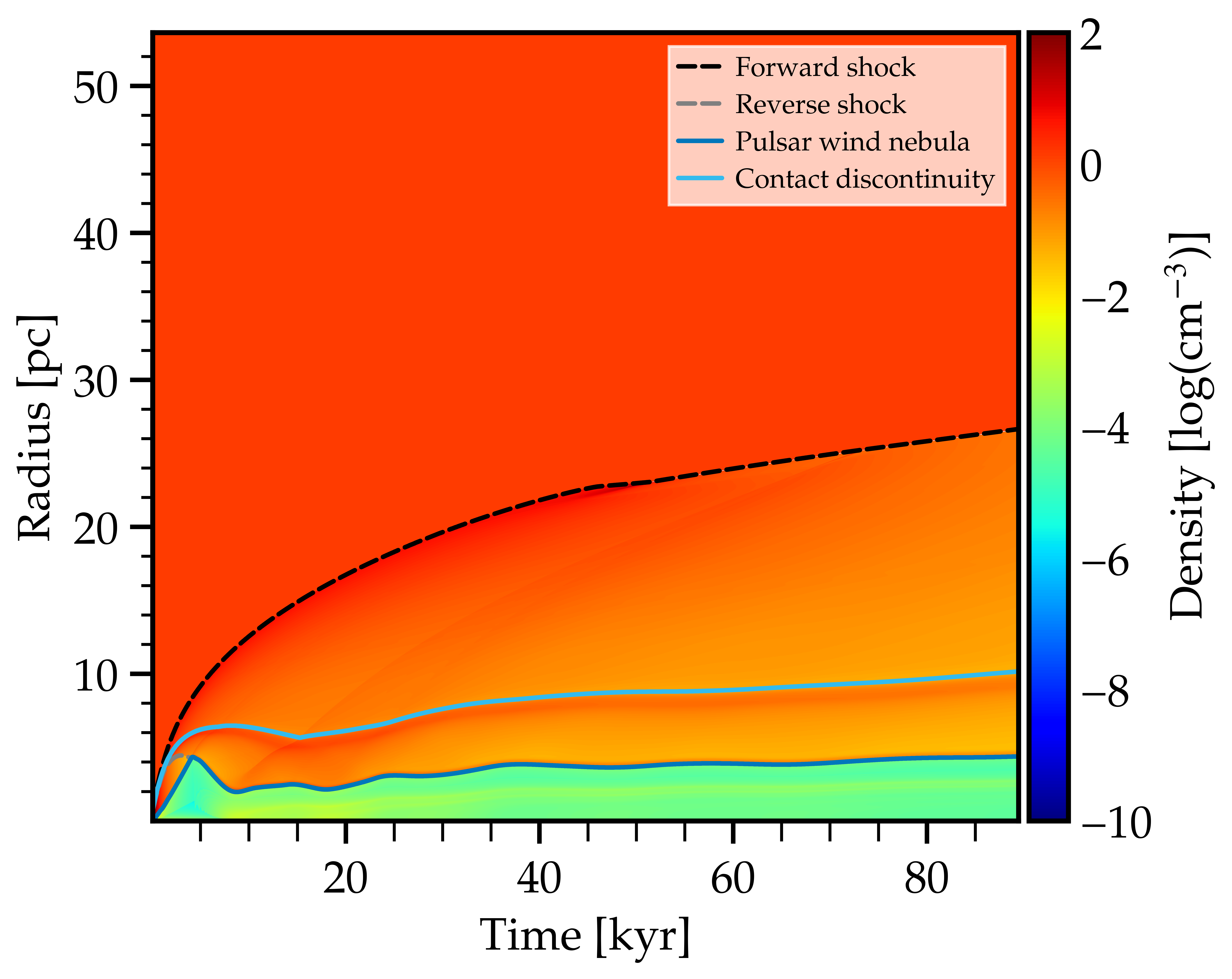}
    \caption{Density kymograph of a PWN+SNR system evolving in the uniform ISM, for the pulsar model L37.5T3.6, born from a M8 progenitor (top panel) or a M20 progenitor (bottom panel). The simulations were performed over the same duration in characteristic units, and hence the different time axes.}
    \label{im:kymograph_8-20_ISM}
\end{figure}

\begin{figure}
    \centering
    \includegraphics[width=0.49\textwidth]{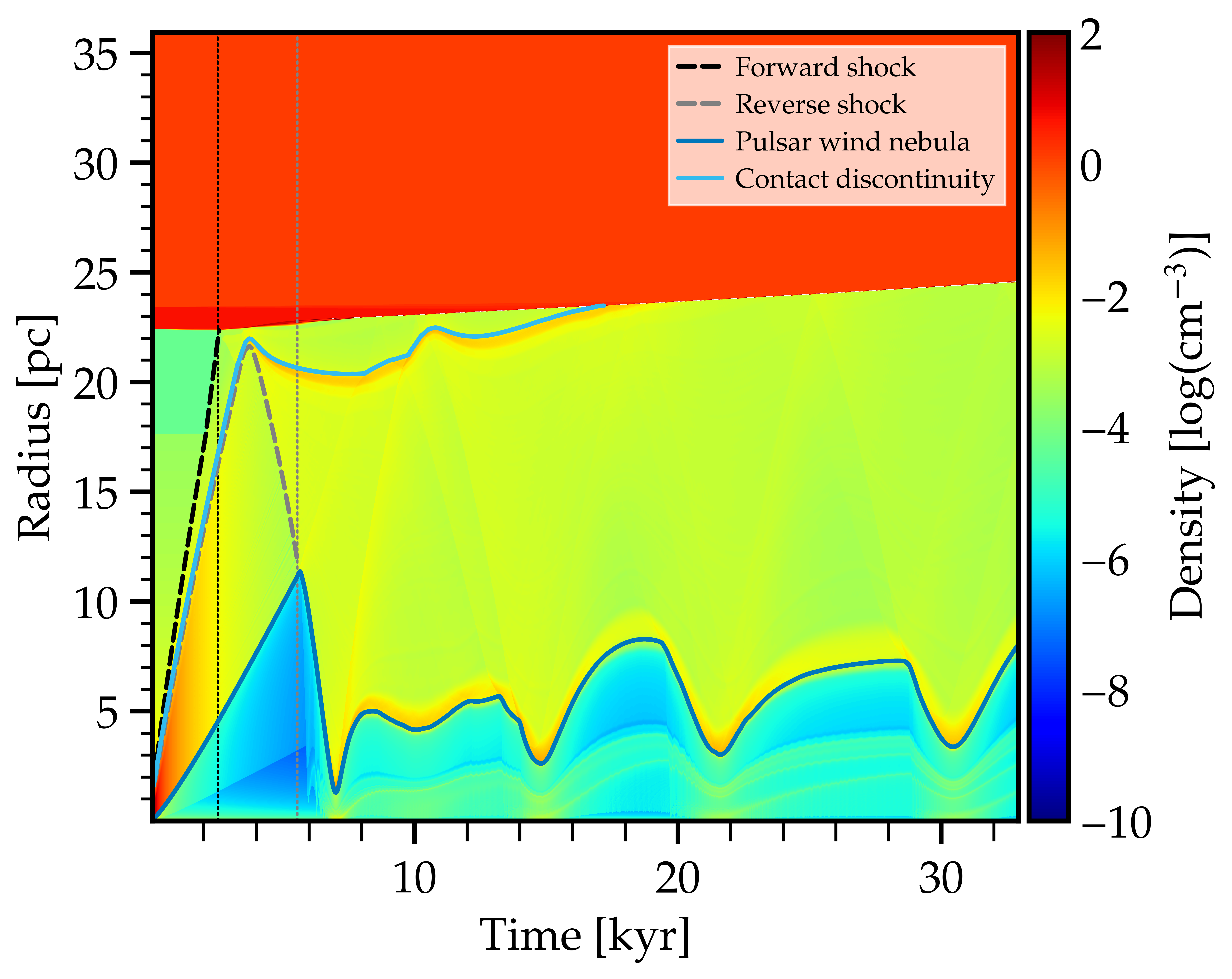}
    \includegraphics[width=0.49\textwidth]{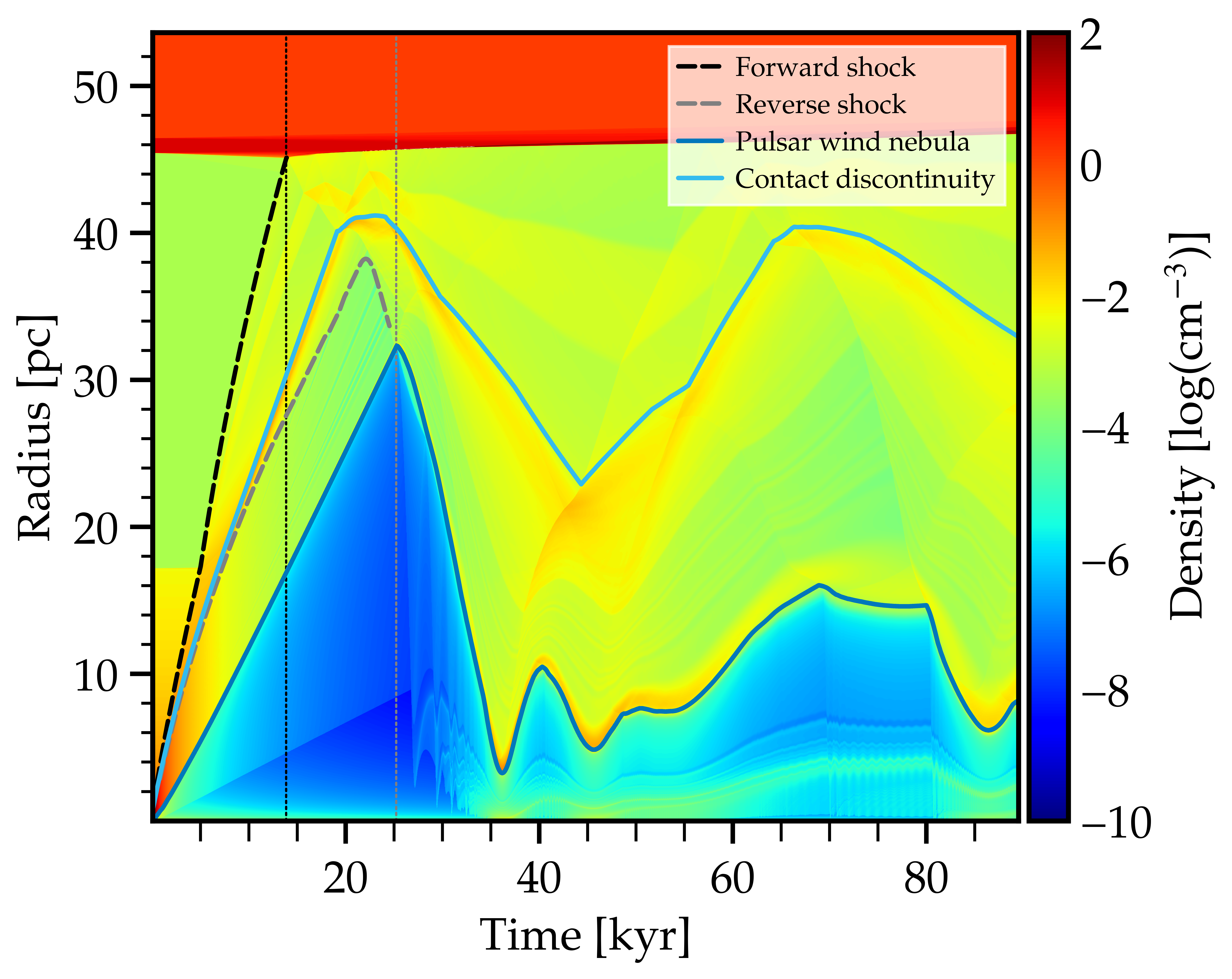}
    \caption{Density kymograph of a PWN+SNR system evolving in a stratified WBB, for the pulsar model L37.5T3.6, born from a M8 progenitor (top panel) or a M20 progenitor (bottom panel). The simulations were performed over the same duration in characteristic units, and hence the different time axes. The black and grey dotted lines are for the FS impact and start of the compression times, respectively.} 
    \label{im:kymograph_8-20_WBB}
\end{figure}

The dynamical evolution of the system is illustrated in the form of space-time diagrams for the main flow quantities, also called kymographs. The evolution of density is plotted in Figs.~\ref{im:kymograph_8-20_ISM}-\ref{im:kymograph_8-20_WBB} for both progenitors and both environments. All four plots share the same colour scale. Overlaid are some important frontiers in the structure of the system: the forward shock of the SNR, in  black; the contact discontinuity (CD) marking the separation between the stellar ejecta and the bubble material ahead of it, in light blue; RS of the SNR, in grey; and the outer boundary of the PWN, in dark blue. A few remarks are in order regarding the WBB cases. First, we stopped tracking the FS after its collision with the WBB shell, mainly for readability purposes, but we emphasise that a transmitted shock propagates across the shell at a much reduced velocity. Second, this same collision produces a strong reflected shock that propagates back inwards. This reflected shock will eventually merge with the reverse shock, which is different in nature, and violently accelerate it. In the following discussion, we shall designate the resulting shock as the reflected shock. In the plots, however, due to difficulties in tracking the reflected shock in the very first stages (complicated patterns of transmitted and reflected shocks, especially in the M20 case), we only outline the up-going reverse shock until it is caught up by the reflected shock, and the down-going reflected shock thereafter (uniquely labeled RS). Last, in the WBB cases, we also highlight some important timescales for the system, namely the time when the FS collides with the WBB shell in black dots and the start of the PWN compression by the RS in grey dots.

In the following subsections, we describe the evolution of the system, starting from the reference ISM case and then addressing the WBB case.
When the same phenomena happen, we give the relevant times and positions of interaction first for the M8 progenitor and then in parentheses for the M20 progenitor.

\begin{figure}
    \centering
    \includegraphics[width=0.49\textwidth]{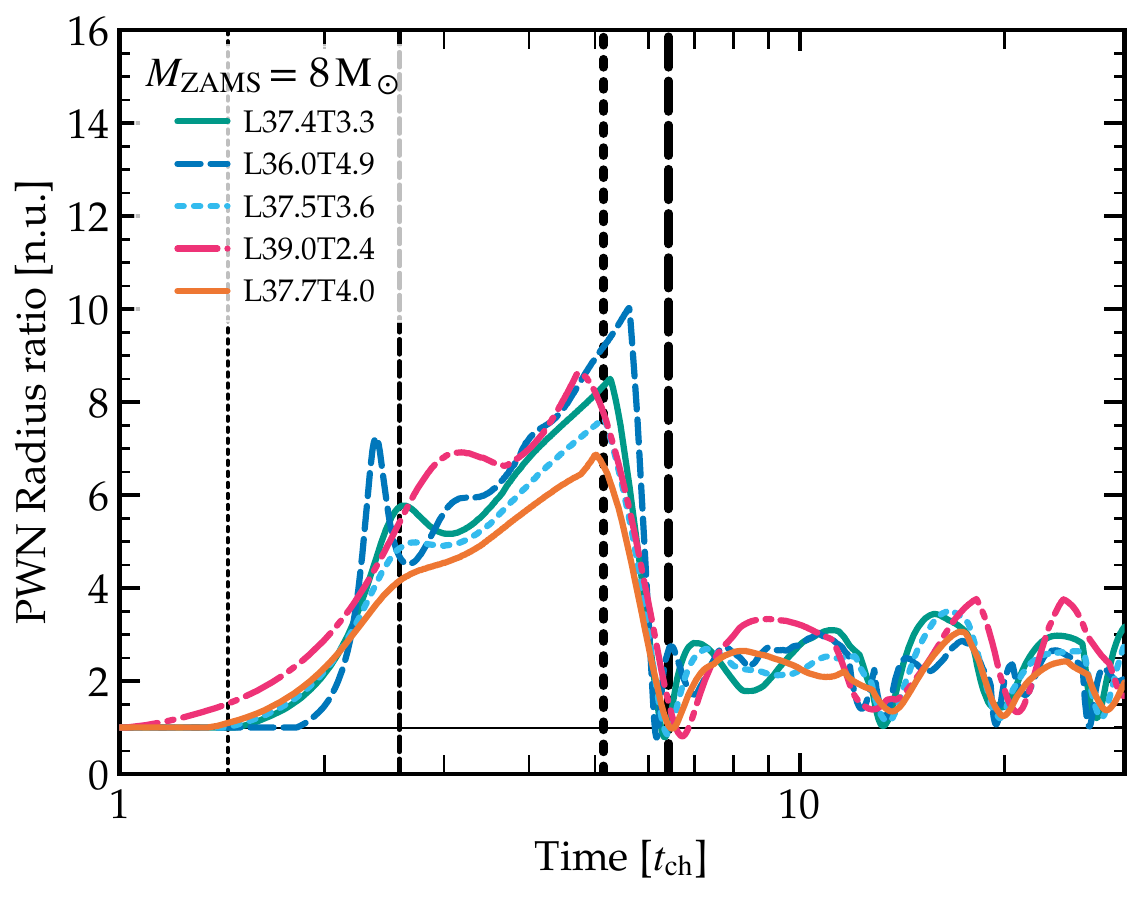}
    \includegraphics[width=0.49\textwidth]{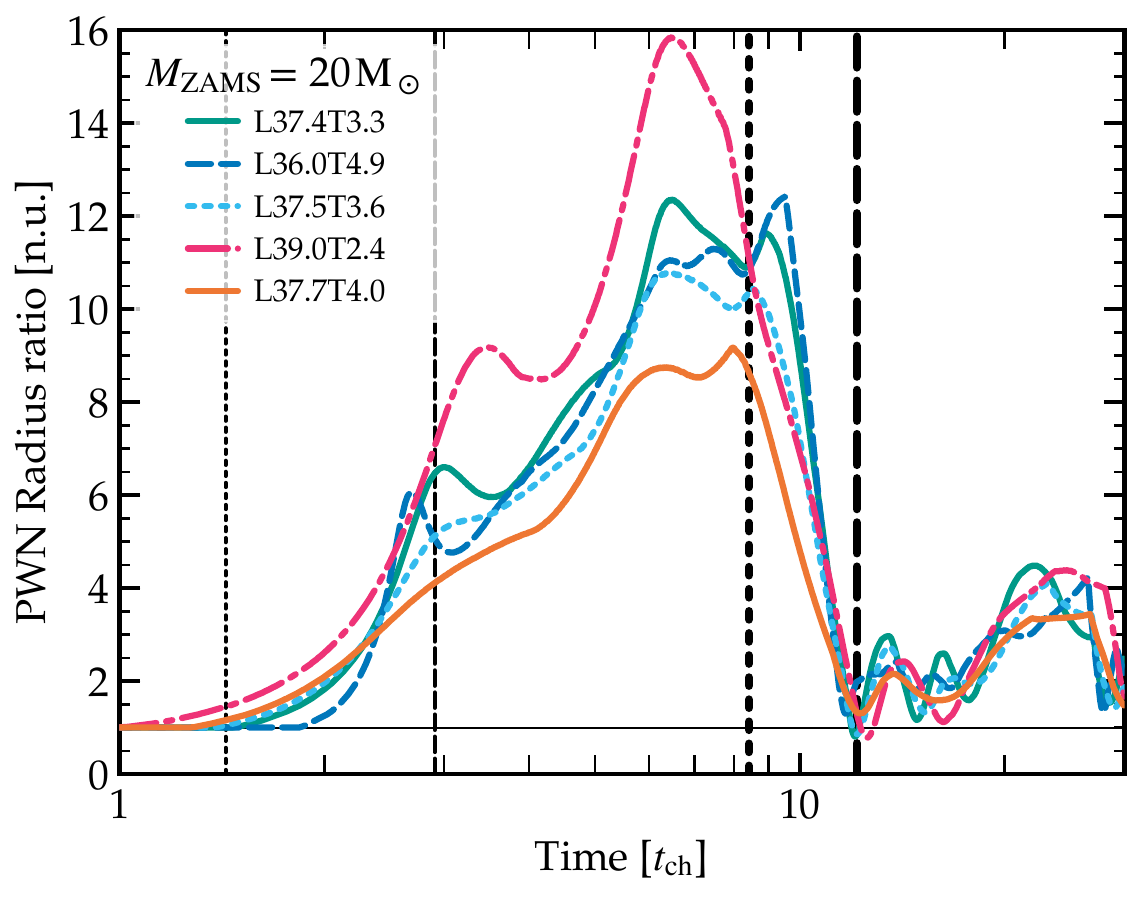}
    \caption{Evolution of the ratio of the PWN radius in the WBB case and in the ISM case, for a M8 progenitor (top panel) and a M20 progenitor (bottom panel). The ratio is shown for all pulsar models in Table \ref{tab:pulsar_parameters}. The dotted lines show the average start time of the compression by a reverse or reflected shock that happens when the PWN reaches its maximum extension. The dashed lines show the average end time of the initial compression and beginning of the reverberation. The thin and thick lines relate to the ISM and WBB cases, respectively.} 
    \label{im:PWN_radius_ratio}
\end{figure}

\subsubsection{PWN+SNR evolving in the uniform ISM}

The two panels of Fig. \ref{im:kymograph_8-20_ISM} display very similar evolutions for both progenitors, with a few noticeable exceptions.
As shown in \citet{truelove+1999}, the evolution of non-radiative SNRs can be described, for a given core+power-law structure of the ejecta, via a unified dimensionless solution connecting periods of self-similar behaviours. 
Expressed in characteristic units, the trajectories of the FS in our simulations are initially identical for both progenitors, after a short transition phase lasting $\sim0.5~t_\mathrm{ch}$ during which the shocks form and are set in motion.
The similarity ends when the M20 system enters the radiative phase at $\sim45~\mathrm{kyr}$, which initiates a period of shallower evolution. The transition time depends only on ejecta energy and ambient density \citep[the information on ejecta mass is lost in the Sedov-Taylor phase as the swept-up mass outweighs it; see][]{blondin+1998}, so it should occur at the same time in the M8 case, beyond the time range simulated here.

The trajectories of the RS in our simulations are also very similar for both progenitors, when expressed in characteristic units, but only up to the moment the shocks start propagating inwards, as could be expected from arguments presented in \citet{truelove+1999} about the structure of the reversed ejecta flow. Past this time, increasing deviations appear, with the RS of the M8 progenitor propagating faster towards the centre. Interestingly, the time of interaction between the RS and the PWN is found to be weakly dependent on ejecta mass in our simulations.\footnote{The evolution of the PWN in free expansion has a weak dependence on ejecta mass \citep[$M_\mathrm{ej}^{1/6}$, when transforming Eq. 12 in dimensionless form in][]{vanderswaluw+2001}, which means that  the PWN is smaller for the lower ejecta mass, which somewhat compensates the effect of a faster down-going RS.} The subsequent evolution of the PWN is specific to each progenitor due to the fact that they reach different sizes before being compressed. As is apparent in Fig. \ref{im:kymograph_8-20_ISM}, the reverberation phases are therefore different, but the evolution eventually converges towards the same late-time trend predicted by \citet{vanderswaluw+2001}, which depends only on the Sedov-Taylor pressure state and pulsar properties (the same for both progenitors).

\subsubsection{PWN+SNR evolving in a stratified WBB}

In contrast to the ISM case, the evolution of the system in a WBB depends strongly on the progenitor, as illustrated in Fig. \ref{im:kymograph_8-20_WBB}. The highly structured density distribution of the environment leads to major qualitative and quantitative differences of the WBB cases from the ISM cases and from each other.

The FS radius initially evolves as $R(t)\propto t^{0.8}$ inside the freely expanding stellar wind, which is expected from the propagation of a blast wave inside a density profile that follows $\rho \propto r^{-2}$ \citep{chevalier1982}.
When entering the bubble interior, the FS first accelerates, owing to the density decrease compared to the freely expanding stellar wind, and then evolves with a nearly constant velocity.
In the M20 case, the ten times larger density and two times larger size of the WBB allows for a progressive transition to a regime with an index similar to that found in the Sedov-Taylor stage, starting at around $5-10~\mathrm{kyr}$.

Eventually, the FS hits the outer shell of the WBB at $2.5~\mathrm{kyr}$ ($13.8~\mathrm{kyr})$, which produces a strong reflected shock propagating back in the interior, and a weaker transmitted shock propagating out across the shell \citep{dwarkadas2005}. The very heavy shell acts nearly like a hard wall such that the velocity of the reflected shock is comparable to that of the incoming FS, while that of the transmitted shock is reduced by about $\sqrt(\rho_\mathrm{shell}/\rho_\mathrm{b})$ (to a factor of order unity in both cases). Here, this means a transmitted shock velocity a few hundred times smaller than that of the impinging FS, and a shell-interior interface set in motion at a fraction of this speed. 

On its way towards the centre of the system, the reflected shock interacts first with the contact discontinuity separating bubble interior and stellar wind, and then with the contact discontinuity separating stellar wind and ejecta, at $3~\mathrm{kyr}$ ($18~\mathrm{kyr}$). This triggers the development of a complex network of secondary transmitted and reflected shocks (particularly visible in the M20 case) travelling back and forth between the contact discontinuities and the bubble shell and interacting with each other. This raises the pressure ahead of the CD and ultimately reverses the flow of material originally set in motion by the passage of the FS. The reflected shock, transmitted across the CD, encounters the RS, hardly detached from the CD in the M8 (the SNR being in ejecta-dominated stage) but sweeping up ejecta at a more significant rate in the M20 case (the SNR being in an almost Sedov-Taylor phase). The two shocks merge into a single blast wave propagating fast towards the centre of the WBB (which, according to our previous definition, we  call RS).

Meanwhile, the PWN expanded freely inside the cold ejecta up to a maximal radius of $11~\mathrm{pc}$ ($32~\mathrm{pc}$), a size that is never reached in the ISM case. The dynamics is well described by the formulae of \citet{vanderswaluw+2001}, for a PWN evolving in a SNR in the ejecta-dominated phase. The collision of the fast RS with the PWN at $5~\mathrm{kyr}$ ($25~\mathrm{kyr})$ initiates a violent compression of the nebula. The latter shrinks by a factor $\sim10$ and reaches its minimum size at $7~\mathrm{kyr}$ ($36~\mathrm{kyr}$) until its internal pressure has risen enough to resist further crushing. Compression proceeds very fast, at nearly the speed of the RS, as can be seen from the slope of the trajectories in Fig. \ref{im:kymograph_8-20_WBB}.

When maximum compression is attained, the imploding ejecta pile up at the outer boundary of the PWN and this generates a strong shock propagating outwards that we call a bounce shock (BS). The very powerful BS achieves a nearly complete thermalisation of the ejecta and then drives the evolution of the whole system through successive back-and-forths across the cavity, following reflections on the shell and transmissions across the CD. In the M8 case, this ultimately leads to a complete collapse of the original WBB material against the shell at $17~\mathrm{kyr}$, making way for the ejecta to entirely fill the cavity. Conversely, in the M20 case, the bubble interior is much more massive and cannot be crushed completely (for a same kinetic energy of the SNR). Instead, the system enters a cycle of alternate compressions and expansions of the ejecta and bubble material.

From this point on, the differences between the ISM and WBB cases are striking. While the ejecta never expand beyond $7~\mathrm{pc}$ ($10~\mathrm{pc}$) in the ISM case (over the time frame explored here), they reach out to $25~\mathrm{pc}$ or beyond when the system evolves inside a WBB. The density of the ejecta is much smaller (because of the larger occupied volume), and their temperature is much higher (because of the multiple shocks crossing the cavity). As we  show later, this will have major consequences in terms of X-ray signature. Regarding the PWN, it is clearly smaller in the ISM case, and its size evolves relatively smoothly after the first compression and bounce.
Conversely, the dynamics of the system is much wilder in the WBB than in the ISM. The crushing of the PWN by the reflected shock produced inside the WBB is much more brutal than that resulting from the reverse shock when the SNR develops in the ISM, and the faster implosion of the ejecta yields a much stronger BS when the collapse is halted as the nebula reaches its minimum size. Combined with a tenuous medium and efficient reflection on the massive and stationary shell, this guarantees a more rapid evolution: the PWN from the M8 progenitor experiences four successive compressions in the WBB case, corresponding to as many two-way crossings of the bubble; in the ISM case, the BS from the first compression and bounce hardly reaches the moving outer edge of the SNR over the same period. It is very likely that the more vigorous motions of the CD and the PWN in the WBB case would lead to an efficient mixing of the different media, shocked pulsar wind with stellar ejecta and stellar ejecta with bubble interior, when taking into account hydrodynamical instabilities and inhomogeneities in the flows (by means of multi-dimensional simulations).

\subsubsection{Impact of pulsar properties}

We compare in Fig.~\ref{im:PWN_radius_ratio} the evolution of the PWN radius when the system evolves in a WBB and in the ISM. The evolution is given as a function of time in characteristic units to be able to compare the two progenitors (we remind that $t_\mathrm{ch}\approx1~\mathrm{kyr}$ and $3~\mathrm{kyr}$ for M8 and M20, respectively). The colours correspond to the different pulsars of Table \ref{tab:pulsar_parameters}, and the vertical lines indicate some reference times in the evolution of the system. The dotted lines denote the maximum extension of the PWN and the start time of its compression by a reverse or reflected shock. The dashed lines show the end time of the initial compression and the beginning of reverberation. The thin and thick lines relate to the ISM and WBB cases, respectively. All vertical lines correspond to averages over the different pulsars.
The reference times are very similar for both progenitors when evolving in the ISM, unsurprisingly because they are in Sedov-Taylor stage and share nearly identical properties. In contrast, there is a noticeable difference between progenitors when the system evolves in a WBB, which can be mainly ascribed to the bubble properties (especially size, which more than doubles with the more massive progenitor). 

The different pulsar models have an influence on the size ratio. This is particularly noticeable for the M20 progenitor because its larger WBB makes it possible for the PWN to grow for a longer period of time and to larger size, thereby revealing more conspicuously the effects. The evolution of the ratios can be decomposed into five phases. The curves are initially flat at a value of one, indicating equal radius, because PWNe are in free expansion inside a SNR in ejecta-dominated state, and this is insensitive to the surrounding medium. The ratio then rises as the PWN is compressed by the RS in the ISM case while it keeps growing in the WBB case. The rise manifests sooner for higher spin-down power pulsars because this implies a faster growth and larger size for the PWN, hence an earlier RS crushing.

At the start of the reverberation in the ISM case, the PWN re-expands more or less rapidly depending on the rotational energy still available, while the PWN is still expanding significantly at its original pace in the WBB case. This results in a drop or flattening of the ratio but overall the ratio keeps growing. In this phase, the ratio depends on both the spin-down power and spin-down time of the pulsar. The highest ratios are obtained for model L39.0T2.4 because, having injected almost all of its energy before $t_\mathrm{ch}<1$, its PWN radius stays almost constant after first compression in the ISM case, while it keeps increasing fast in the WBB case. Conversely, model L37.7T4.0 is associated with the smallest ratios because, having the highest rotational energy of all pulsars and a spin-down time of $10~\mathrm{kyr}$, it keeps injecting energy past first compression and its PWN is the one that grows the most in the ISM case. The combination of high spin-down power and long spin-down time yields this result, and the single spin-down time is not enough: pulsar model L36.0T4.9 has a very high long spin-down time but its spin-down power is so low that its PWN remains small in the ISM case.

The ratios increase until the RS-PWN interaction occurs in the WBB case. This is when the highest ratios are achieved, with values as high as 8 (16). The violent compression of the PWN in the WBB case causes a strong reduction of the ratio, albeit never or hardly below one, until the reverberation of the PWN in the WBB starts. In the subsequent phase, the PWN in the WBB is always bigger than that inside of the ISM, with a size oscillating from roughly the same size up to larger by a factor of a few. The size ratio is always greater than one because the pressure in the surrounding ejecta is smaller by an order of magnitude in the WBB case compared to the ISM case (the temperature is an order of magnitude higher, but the density is two orders of magnitude lower because the ejecta could expand much more). The late evolution displays some similarities despite the very different pulsar properties, and this can be ascribed to the reflected shock, which drives most of the reverberation dynamics and is entirely determined by the SNR and WBB properties for a given progenitor (see the previous section).

We further comment here on our assumption of a 1D model with a central static pulsar, with zero kick velocity. In reality isolated neutron stars gain a kick velocity whose distribution peaks at $200~\mathrm{km/s}$ and has a median at $400~\mathrm{km/s}$ \citep{disberg+2025} due to the asymmetries in the SN explosion and/or ejecta fallback on the neutron star. Assuming that the pulsar has a natal kick velocity of $400~\mathrm{km/s}$, the time at which it definitively escapes the PWN is about $3~\mathrm{kyr}$ for M8 and $7~\mathrm{kyr}$ for M20 in the ISM case, and about $20~\mathrm{kyr}$ for M8 and $35~\mathrm{kyr}$ for M20 in the WBB case (for pulsar model L37.5T3.6). Spin-down timescales typically being in the $10^2-10^4$\,yr range, the larger size of a PWN in a WBB environment will allow most pulsars to release the most of their rotational energy inside the nebula. Conversely, a smaller PWN in an ISM environment makes it possible for the pulsar to escape into the SNR while still being very powerful. This has a number of implications especially in terms of non-thermal radiative signature: PWNe developing in WBB environments will show up as one single very extended emission component, while PWNe in a normal ISM environments may appear as composite sources comprising both a relic and a newborn nebula of smaller sizes and possibly comparable intensities. The latter set-up has to be investigated by means of 2D or 3D simulations, including other effects like anisotropies or inhomogeneities in the environment \citep{kolb+2017}.

\begin{figure}
    \centering
    \includegraphics[width=\columnwidth]{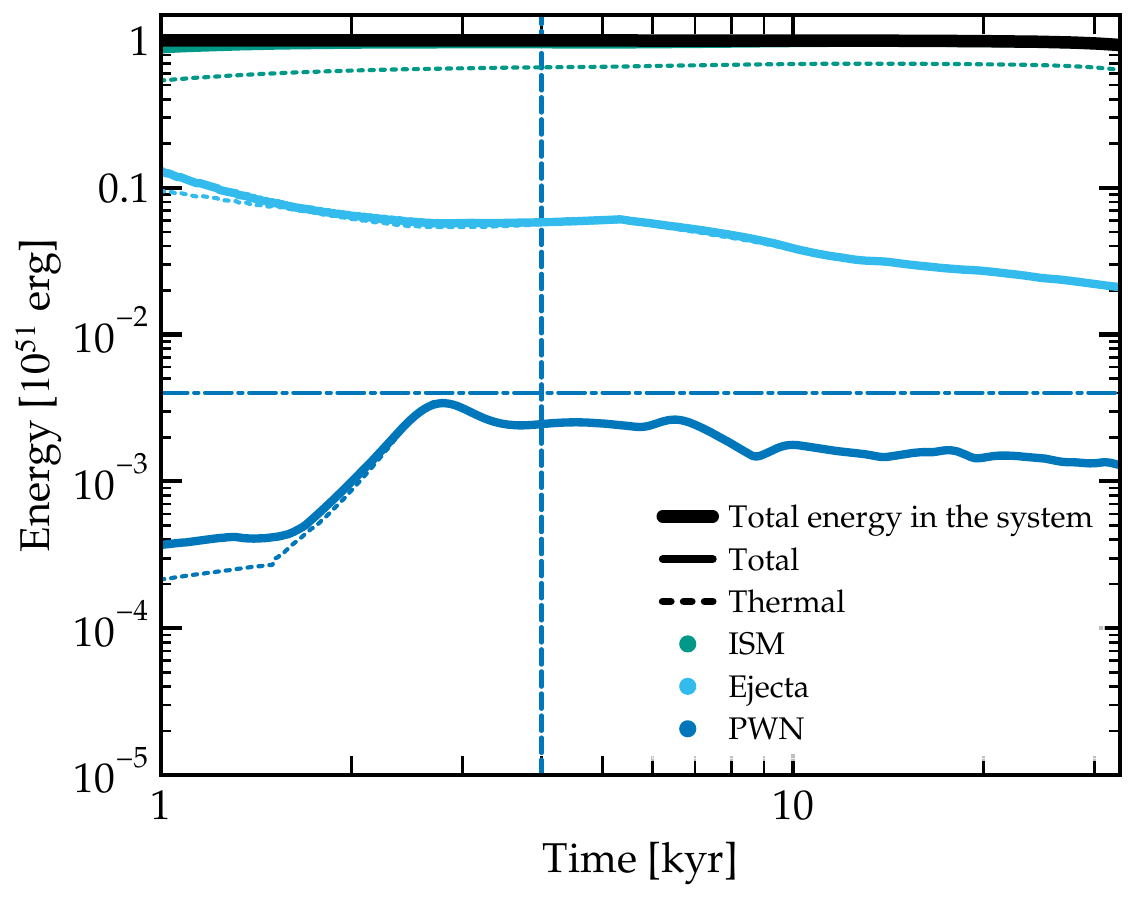}
    \includegraphics[width=\columnwidth]{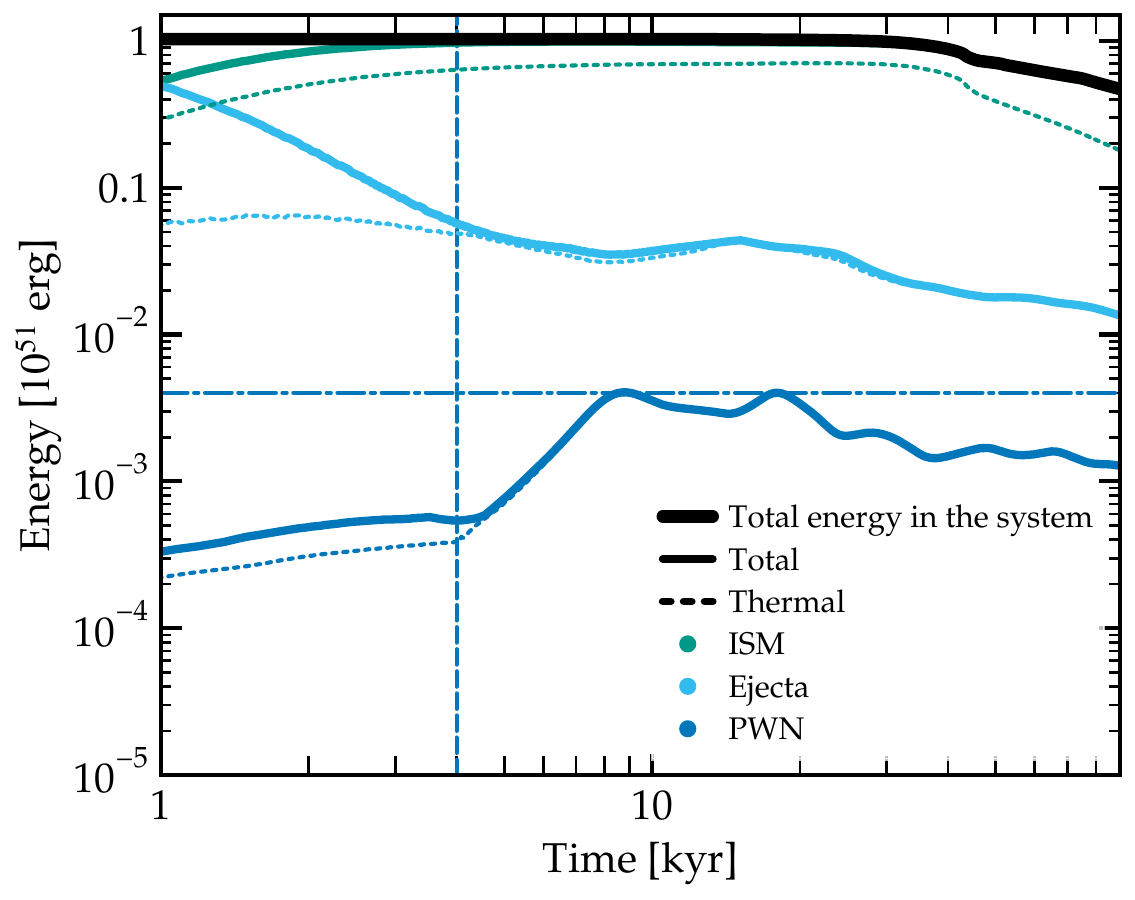}
    \caption{Time evolution of the energy in each medium, for the simulations involving pulsar model L37.5T3.6, a PWN+SNR system evolving in the uniform ISM, and a M8 or M20 progenitor (top and bottom panel, respectively). We highlight the total rotational energy and pulsar spin-down time as a blue horizontal dash-dotted line and a blue vertical dashed  line, respectively.}
    \label{im:energy_budget_ISM}
\end{figure}

\begin{figure}
    \centering
    \includegraphics[width=\columnwidth]{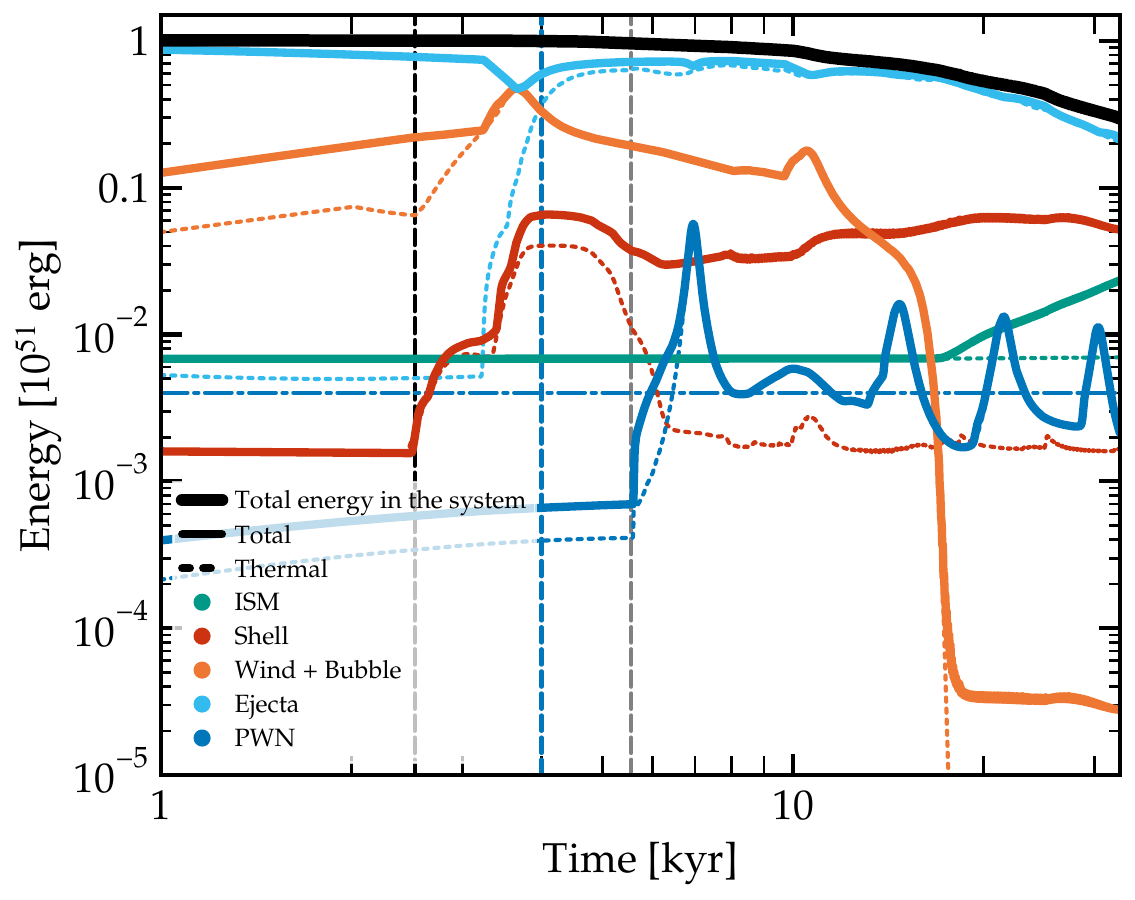}
    \includegraphics[width=\columnwidth]{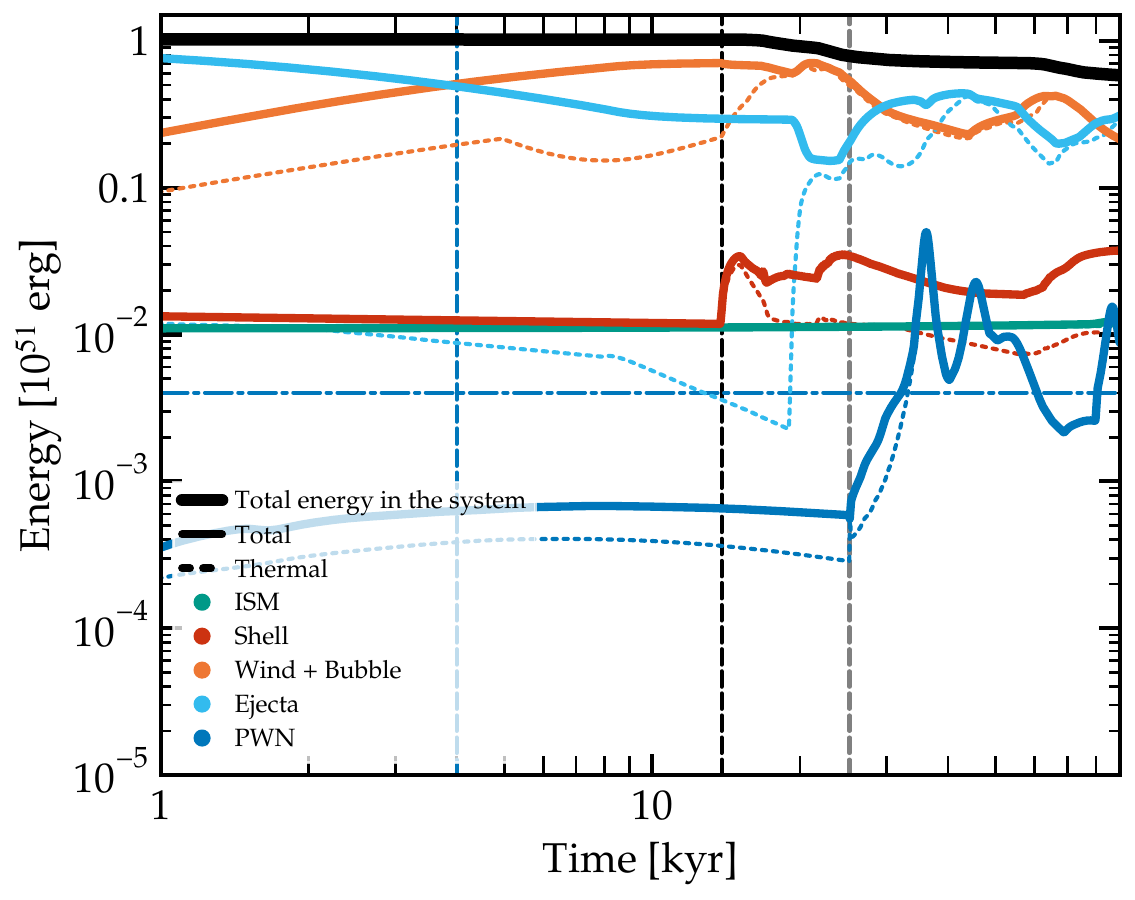}
    \caption{Time evolution of the energy in each medium, for the simulations involving pulsar model L37.5T3.6, a PWN+SNR system evolving in a stratified WBB, and a M8 or M20 progenitor (top and bottom panel, respectively). We highlight the total rotational energy and pulsar spin-down time as a blue horizontal dash-dotted line and a blue vertical dashed  line, respectively. The black and grey dashed lines represent the FS impact and start of the compression times, respectively.}
    \label{im:energy_budget_WBB}
\end{figure}

\subsection{Energetics}

In order to give a complementary perspective on the evolution of the PWN+SNR system, we computed the thermal and kinetic energy in each medium composing it, as a function of time. This is presented in Fig. \ref{im:energy_budget_ISM} for the ISM case, and in Fig. \ref{im:energy_budget_WBB} for the WBB case, for both stellar progenitors and our reference pulsar model. The thermal energy is given as dotted lines of different colours for the different media, and the total energy as solid lines (the kinetic energy is the difference of the latter two). For comparison, we highlight the spin-down time of the pulsar as well as its total rotational energy. In the WBB case, we separated the ISM swept-up in the bubble shell from that still at rest outside of it, and we grouped the initially freely expanding stellar wind and bubble interior material into a single component.

\subsubsection{PWN+SNR evolving in the uniform ISM}

Over the first $1~\mathrm{kyr}$ ($3~\mathrm{kyr}$), the SNR transfers most of its original kinetic energy to ISM material, and the ejecta is almost entirely thermalised by the RS at $2~\mathrm{kyr}$ ($5~\mathrm{kyr}$). 
The swept-up ISM component dominates the energetics of the system by far during the rest of the simulation, and starts to significantly lose energy by thermal radiation at $\sim 30-40~\mathrm{kyr}$.
The ejecta almost continuously lose energy to the expansion of the SNR and the compression of the PWN (and to a much lesser extent by radiation in the region of the CD), except a modest increase following the maximum shrinking of the PWN (the first peak in the dark blue curves): the bounce shock generated by imploding ejecta being suddenly halted restores some energy to the SNR over a few kiloyears.

The PWN gradually gains energy during compression by the reverse shock, until its minimum size is attained. At this point, its energy is predominantly thermal and amounts to the total energy released by the pulsar. The PWN then follows a global trend of continuous energy loss as it keeps growing in size (with modest up and down fluctuations due to reverberation), to eventually stabilises at $10^{48}~\mathrm{erg}$ for both progenitors. 

\subsubsection{PWN+SNR evolving in a stratified WBB}

A noticeable difference between the WBB and ISM cases is that the ejecta retain their energy for a longer time when the system evolves in a bubble. This is first due to the low densities, which delay the transfer of energy to the surrounding medium, and then to the massive bubble shell, which acts as a reflecting wall and ensures that most of the energy remains flowing inside. At the end of our simulations, with either progenitor, the ejecta retain several tens of per cent of their original $10^{51}~\mathrm{erg}$ energy budget ($30\%$ at $30~\mathrm{kyr}$ for a M8 star and $50\%$ after $90~\mathrm{kyr}$ for a M20 star), to be compared to about one per cent in the ISM case.

The behaviour however differs between progenitors, following in that the respective dynamics discussed in Sect.~\ref{sec:dynamics}. 
Over the first few kiloyears, the ejecta of a M8 progenitor only start transferring their kinetic energy to the bubble material, while the denser bubble interior of a M20 progenitor leads to more rapid conversion to the point that the bubble material dominate the energetics at $5~\mathrm{kyr}$. When the FS collides with the bubble shell, the strong reflected shock first slows down and thermalises the shocked bubble material, then crosses the CD, and slows down and thermalises the shocked ejecta (actually, one should talk about a network of reflected and transmitted shocks instead of a single reflected shock; see Sect.~\ref{sec:dynamics}). This causes the concomitant dip and jump in the light blue and orange curves, respectively, and the strong rise in the light blue dotted lines, in Fig. \ref{im:energy_budget_WBB}. The first peak in the energy of the bubble material is due to its maximum squeezing between the ejecta and the shell. From then on, the re-expansion of the bubble material towards the interior causes its energy to decrease, while that of the ejecta increases anew. The subsequent evolution is markedly different for both progenitors. In the M8 case, the strong bounce shock produced after the maximum compression of the PWN (at $7~\mathrm{kyr}$) will travel back and forth in the cavity, heat up the ejecta, and eventually crush the bubble material to the point that it merges completely with the shell at $17~\mathrm{kyr}$. Conversely, in the M20 case, the more massive bubble interior prevents its total collapse and, instead, the energies of the ejecta and bubble material evolve out of phase, reflecting the cycles of alternate compressions and expansions.

Meanwhile, the bubble shell energy increases as a result of successive collisions with blast waves, which generates low-velocity transmitted shocks across it. The initial jump is caused by the impact of the FS, and the subsequent increases are due to the multitude of secondary up-going reflected shocks. Most of the shell energy is communicated when the bubble material undergoes its first compression against the shell. While the FS impact is the most significant contribution for the M20 case, secondary shocks are dominant in the M8 case. The shell actually is the main site of thermal radiation losses, as a result of the low-velocity transmitted shocks. This is reflected in the decreases of its thermal energy content (the red dotted lines). The thermal energy tends to naturally return to the original value, before the shock interactions, because it is a thermally stable state where losses are negligible.
Radiative shocks transmitted in the shell eventually cross it entirely, enter the undisturbed ISM and put it in motion. This causes an increase in the kinetic energy of the latter, at $18~\mathrm{kyr}$ ($80~\mathrm{kyr}$).
It is interesting to note that the total energy lost to radiation is much higher for a M8 progenitor evolving in a WBB compared to the ISM case (about 70\% versus less than 10\%). As illustrated in Fig. \ref{im:energy_budget_WBB}, this can be ascribed to two main causes. First, the collapse of the bubble material, which is a major energy component. Because of its very low density and low mass, the bubble material can be completely fused together with the shell material, which allows its energy to be radiated away. Second, still because of its small mass, the bubble material is a much less effective buffer zone between the ejecta and the shell. As a result, much more energy can be passed from the former to the latter, as can be seen from the solid red curves in Fig. \ref{im:energy_budget_WBB}: they reached significantly higher levels in the M8 case, nearly up to 10\% of the original ejecta energy. For the M20 progenitor, the energy lost to radiation is comparable in the WBB and ISM cases, but this is mostly chance coincidence as radiation sites are located in different places in space and time.

At the centre of the system, the energy in the PWN is the same for both progenitors in the early stages, when it is freely expanding in the cold ejecta (apart from minor differences at times $\lesssim 0.5 t_\mathrm{ch}$ when the flow progressively sets up from the initial state). The PWN reaches a first maximum in energy after about two pulsar spin-down times. Slightly before or after that, depending on the progenitor, the collision of the expanding PWN with the down-going RS initiates the compression phase (highlighted as vertical grey dashed lines). This translates into a dramatic increase of the PWN energy, by roughly two orders of magnitude, up to when the minimum size of the PWN is reached. At this time, the energy is mostly thermal and exceeds the total rotational energy of the pulsar by about an order of magnitude. In our simulations, the PWN is a very hot medium that cannot radiate energy away (its temperature exceeds $10^8$\,K, the high end of our cooling function). Had we implemented non-thermal components, the magnetic fields would have been compressed in the PWN, drastically increasing their strength and burning off the relativistic lepton component via synchrotron radiation. The re-expansion of the PWN causes its energy to decrease and the subsequent reverberation phase translates into up and down fluctuations of the energy, inversely proportional to the size of the nebula. From this point on, a key difference with the ISM case is that the PWN energy exceeds the total pulsar rotational energy for most of the time, by factors of a few and up to about ten. In a WBB environment and at late stages, most of the PWN energy comes from the remnant and not from the pulsar. This is due to the strong reflected shock produced when the SNR collides with the massive bubble shell, which is much more efficient at returning ejecta energy to the PWN than the milder reverse shock in the ISM case.

\subsection{X-ray emissions}

\begin{figure}
    \centering
    \includegraphics[width=\columnwidth]{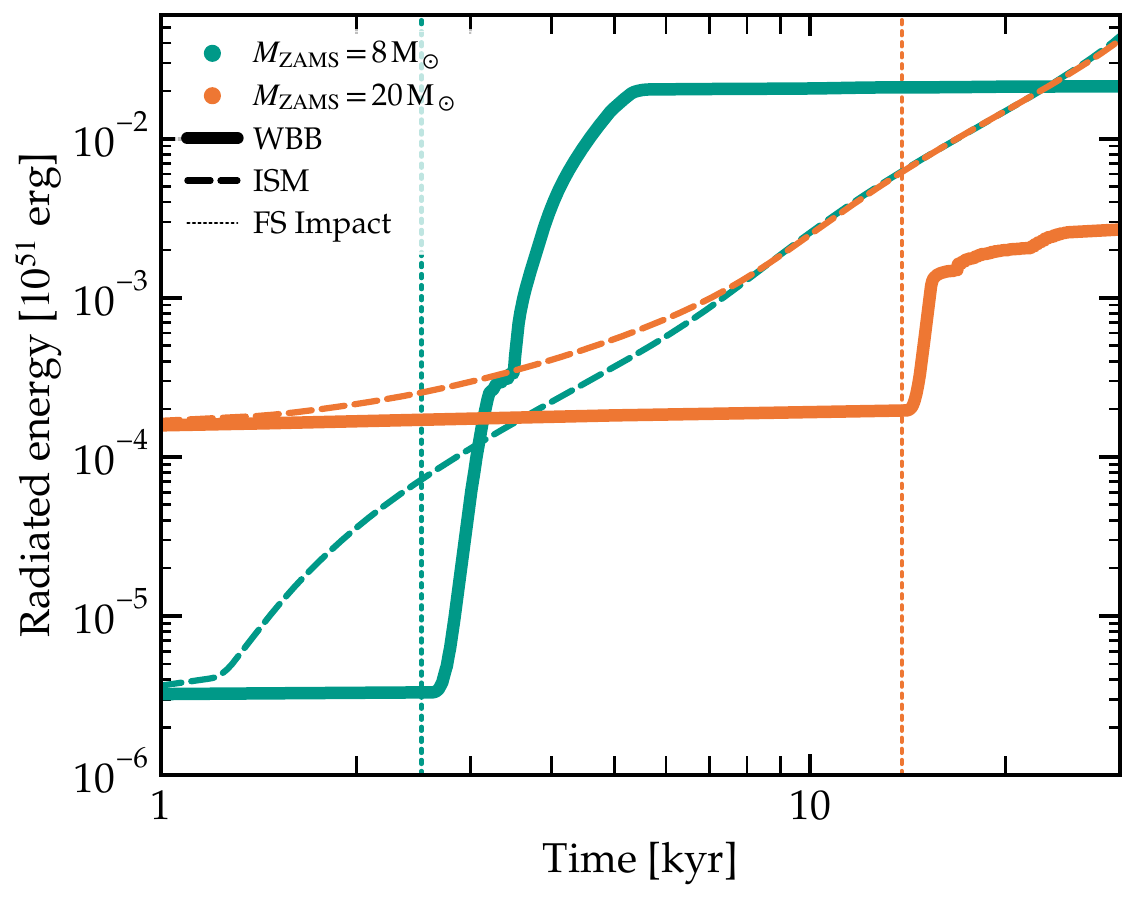}
    \includegraphics[width=\columnwidth]{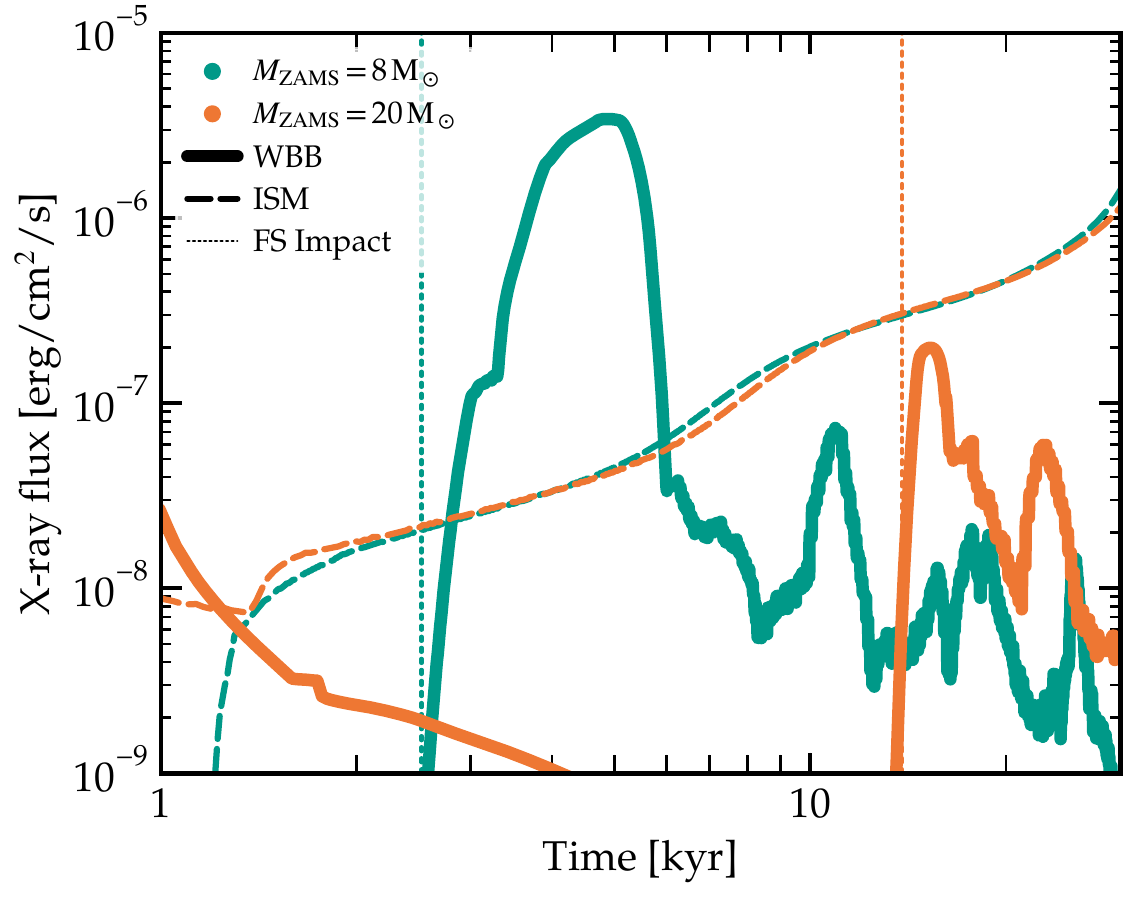}
    \caption{Cumulative energy and instantaneous flux (top and bottom panels, respectively) radiated in the $0.2-10.0~\mathrm{keV}$ band, as a function of progenitor mass and environment. We mark as vertical dotted lines the time at which the FS collides with the bubble shell in the WBB cases.}
    \label{im:radiated_energy}
\end{figure}

Based on our hydrodynamical simulations, we computed the associated thermal emission, especially in the X-ray range, which is a window of major interest for the study of SNRs \citep{vink2012}. We used XSPEC \citep{XSPEC}, an X-ray spectral-fitting and emission-modelling code, to compute the X-ray emission spectrum from each cell in our grid, based on the predicted temperature and density in the cell. We assumed an unabsorbed APEC emission model and a solar abundance. The latter hypothesis may not be fully appropriate for emission from a metal-rich ejecta, but this turns out to be a minor contribution to the emission, which is instead dominated by shocked ISM. We are primarily interested in comparing the WBB and ISM cases in terms of their X-ray flux in a typical band for current telescopes like the Chandra X-ray Observatory, XMM-Newton, or eROSITA. We therefore integrate all spectra over the $0.2-10.0~\mathrm{keV}$ photon energy band and summed the emission from all cells in the grid.
We show the cumulative energy lost as thermal radiation in the $0.2-10.0~\mathrm{keV}$ energy band as a function of time in Fig.~\ref{im:radiated_energy}, for both progenitors and both environments, and our reference pulsar model.

When the PWN+SNR system is evolving in the uniform ISM, the X-ray emission differs only in the first few thousands of years for the two progenitors. The different ejecta masses lead to different density and temperature structures in the CD region, which is where most of the radiation comes from in the early stages. Once in the Sedov-Taylor stage, however, both SNRs behave identically and so does their emission. The main emission site in this phase is the region downstream of the FS (the CD region keeps radiating but at a more modest level).

When the PWN+SNR system is evolving in a stratified WBB, the radiation history is very different. The emission remains at a lower level for a longer time. In the M8 case, the bubble material has very low densities and a low mass, the SNR remains in ejecta-dominated stage throughout, with a nearly constant high shock speed yielding very high post-shock plasma temperatures, hence low emissivities. In the M20 case, the higher density and mass of the bubble interior causes the SNR to start its transition towards the Sedov-Taylor stage, during which shocked material begins to radiate as the forward blast wave slows down and shocked ejecta piles up behind the CD creating a density enhancement (the radiated energy is however insignificant for the dynamics of the system at this point). 

The collision of the FS with the bubble shell causes a rapid rise in the emission. The energy transferred to the shell via transmitted shocks is rapidly radiated away within a few kiloyears, mainly at energies below $0.2~\mathrm{keV}$ because of the low temperatures downstream of the low-velocity transmitted shocks. This is achieved in multiple steps (two of which are clearly visible for the M8 case in Fig. \ref{im:radiated_energy}): a first increase follows the impact of the FS on the shell at $2.5~\mathrm{kyr}$ ($13~\mathrm{kyr}$), which creates both a transmitted shock and a strong reflected shock; subsequent emission increases occur as a result of secondary up-going reflected shocks produced as the original reflected shock interacts with discontinuities on its way to the centre. In both cases, one may notice an additional small emission rise at around $10~\mathrm{kyr}$ ($22~\mathrm{kyr}$), more easily seen in the flux plot (bottom panel), caused by the BS crossing the CD. Overall, after the first emission peak, thermal radiation follows a long-term decreasing trend, with some bursts of emissions when secondary or tertiary shocks reach the shell. Such a behaviour is similar to that presented in \citet{krause+2014} for SNRs evolving in superbubbles.

In the end, the evolution of the M8 progenitor yields a higher total radiation output (than the M20 case), because its SNR retains a maximally powerful blast wave until shell collision, with less energy dissipated in the bubble material. But the important point is that, compared to the ISM case, systems evolving in a WBB are characterised by a smaller or much smaller total radiative output in the long run, produced from very localised emission bursts. The probability of observing such a system in X-rays is therefore strongly diminished, and limited to young systems in the case of low-mass progenitors (our M8 case).

We emphasise, however, that the development of a SNR-PWN system in a WBB can be probed in other ways. Optical line signatures in particular are well suited to reveal the transmitted shocks crossing the bubble shell at low to moderate velocities (depending on the actual density structure of the shell). Recent applications to SNRs suspected to be developing inside WBBs can be found in \citet{vogt+2017,payli+2024,bakis+2025}.

Last, we caution that the above results on the emission and energetics may be best considered as qualitative rather than quantitative. As the bubble shell appears as a key location for energy dissipation in the form of radiation, the results are expected to be sensitive to assumptions regarding its structure. In our simulations, we assumed a geometrically thin shell with a fixed width of $1~\mathrm{pc}$ (a few per cent of its radius). Although not tested in the context of this work, a thicker shell would impact the velocity of transmitted shocks and the emissivity of the shocked shell material, which would in turn translate into different intensities and spectra for the thermal radiation.

\section{Conclusions}
\label{sec:conclusions}

For this work we performed a series of numerical experiments to study how a stratified WBB environment affects the development of a PWN and its parent SNR. We built a 1D hydrodynamical analogue of the whole system and ran numerical simulations with the Idefix code. We considered two stellar progenitors with $8~\mathrm{M}_\odot$ and $20~\mathrm{M}_\odot$ initial masses, and five pulsar models with different initial powers and spin-down times representative of the core of the Galactic population. We compared our findings to the reference case of a system evolving in a uniform typical ISM in terms of dynamics, energetics, and X-ray thermal emission.

The WBB environment extends over a 20 to $50~\mathrm{pc}$ radius (larger size for a higher-mass progenitor) and is characterised by densities that are orders of magnitude lower than the typical quiescent ISM. It is bounded by a massive thin shell of $\sim10^3-10^4~\mathrm{M}_\odot$ of interstellar material, which  acts nearly as a hard outer wall. The latter is instrumental in returning energy to the inner parts of the system. 

Because of the low interior densities in WBBs, SNRs can initially grow to much larger sizes and their transition to the Sedov-Taylor stage is delayed. As a consequence, stellar ejecta can extend about $5-6$ times further out. The impact of the blast wave on the bubble shell after a few kiloyears generates strong shocks reflected off the shell, bouncing back at the center, and subsequently travelling back and forth in the cavity. These efficiently thermalise the stellar ejecta, which ultimately retains a sizeable fraction of the initial supernova energy (a few tens of  per cent compared to one per cent for a system in the ISM). Overall, the material inside the bubble is characterised by very low densities and very high temperatures, and hence low thermal emissivities. Most of the thermal radiation is produced by low-velocity transmitted shocks crossing the bubble shell when the shell is hit by a blast wave. Eventually, the total radiative output of systems evolving in a WBB is smaller than for those evolving in the ISM, and most importantly comes from a few very localised emission bursts. This, and their larger size, very likely hinders their detection in thermal X-rays.

The low-density interior of WBBs also allows the PWN to grow to considerably larger sizes during its free expansion stage, more than $30~\mathrm{pc}$ for the $M=20~\mathrm{M}_\odot$ progenitor. The interaction with the strong down-going reflected shock leads to a violent compression in size by an order of magnitude. This is followed by a wild reverberation phase driven by rapid back-and-forth reflections inside the cavity, with successive changes in the size of the PWN by factors of a few. These dynamics may lead to very efficient mixing in multidimensional simulations. At late times, a PWN in a WBB environment is $2-4$ times larger than one in the ISM, on average, and contains far more energy: the latter exceeds that injected by the pulsar, by up to a factor ten, and was extracted from the SNR through the strong reflected shocks. We find that the relative properties of a PWN inside a WBB environment are weakly dependent on pulsar properties, especially during the reverberation phase, but depend significantly on the bubble properties (which are directly linked to the progenitor mass).

Our work provides a possible explanation for the observed large sizes of many PWNe and the frequent non-detection of their parent SNRs. Further investigations of this scenario will benefit from continued observations of SNR-PWN systems with large field-of-view X-ray telescopes such as eROSITA \citep{khabibullin+2024} or Einstein Probe \citep{chi+2026}, and from radio surveys with instruments sensitive to intermediate angular scales such as ASKAP and MeerKAT. On the theoretical side, a natural extension of this work is to evolve towards multidimensional simulations to incorporate the effects of an offset supernova explosion in the cavity, a non-zero pulsar kick, the possibility of efficient mixing through instabilities, and the impact of inhomogeneities and anisotropies in  matter distribution. It is also desirable to evaluate the impact of varying the outer density, and in particular considering higher values than used here, of the order of $\sim10^{2}~\mathrm{cm}^{-3}$ and more characteristic of the dense phases of the ISM where massive stars form \citep{chiotellis+2024,chevalier+1989}.

\begin{acknowledgements}
This work has made use of NASA's Astrophysics Data System Bibliographic Services.
The authors thank the referee for their questions, that greatly participated in improving the quality of the manuscript.
Both authors thank Zakaria Meliani and Niccol\`o Bucciantini for their advice on hydrodynamical simulations.
Lioni-Moana Bourguinat acknowledges the hospitality and support of IRAP, where part of this work was carried out. 
This work was granted access to the HPC resources of CALMIP supercomputing centre under the allocation P23022.
Pierrick Martin acknowledges financial support by ANR through the GAMALO project under reference ANR-19-CE31-0014.
\end{acknowledgements}

\bibliographystyle{aa}
\bibliography{Bibliography/biblioPWNe.bib}

\end{document}